\documentclass[twocolumn, aps, amssymb, prl, superscriptaddress]{revtex4-1} 
\usepackage{amsmath, amssymb, amsthm, wasysym}

\usepackage[section]{placeins}

\usepackage{rotating}
\usepackage{floatpag}
\rotfloatpagestyle{empty}
\usepackage{graphicx}
\usepackage{multind}
\usepackage{hyperref}
\usepackage{epstopdf}

\begin{document}

\title{Cavity quantum electrodynamics with three-dimensional photonic bandgap crystals}

\author{Willem L. Vos}\email{w.l.vos@utwente.nl; www.photonicbandgaps.com}
\affiliation{Complex Photonic Systems (COPS), MESA+ Institute for Nanotechnology, University of Twente, 7500 AE Enschede, The Netherlands}
\author{L\'eon A. Woldering}\email{l.a.woldering@utwente.nl}
\affiliation{Transducer Science and Technology (TST), MESA+ Institute for Nanotechnology, University of Twente, 7500 AE Enschede, The Netherlands}

\date{Prepared on May 21st, 2013, corrected October 26th, 2014. \\ .. From: \emph{Light Localisation and Lasing}, Eds. M. Ghulinyan, L. Pavesi, Cambridge Univ. Press (2015) Ch. 8, p. 180}

\begin{abstract}
Three-dimensional (3D) photonic crystals with a 3D photonic bandgap play a fundamental role in cavity quantum electrodynamics (QED), especially in phenomena where the local density of optical states is essential.
We first review the current status of the fabrication of 3D photonic crystals with a bandgap at optical frequencies, corresponding to wavelengths below 2500 nm.
We discuss the main implications of 3D bandgaps for cavity QED, in particular spontaneous emission inhibition of emitters embedded in a 3D bandgap crystal.
Moreover, we review progress in enhanced emission of emitters placed in a photonic bandgap cavity, thresholdless laser action in a miniature photonic crystal cavity, breaking of the weak-couping limit of cavity QED.
Finally, we discuss several exciting applications of 3D photonic band gap crystals, namely the shielding of decoherence for quantum information science, the manipulation of multiple coupled emitters including resonant energy transfer, lighting, and possible spin-off to 3D nanofabrication for future high-end computing.
\end{abstract}
\maketitle

\section*{Introduction}
The propagation of light in periodically ordered photonic crystals bears a strong analogy to the wave propagation of conduction electrons in atomic crystals.
The development of Bloch modes and the dispersion are determined by the interference of waves that are diffracted by lattice planes~\cite{Bloch1929ZP, Ashcroft1976}.
The periodicity commensurate with the wavelength ($a \approx \lambda/2$) gives rise to Bragg diffraction that is associated with frequency windows for which waves are forbidden to propagate in a particular direction.
Such \emph{stop gaps} have long been known to arise for electromagnetic waves, notably for X-rays in atomic crystals~\cite{James1954}.
At optical frequencies, the stop gaps that occur in one dimensional periodic structures, known as Bragg stacks, are widely used in optical laboratories as broadband highly reflecting dielectric mirrors~\cite{Yariv1983}.

The distinguishing feature of three-dimensional (3D) photonic crystals is that a common stop gap can be achieved for all directions and for all polarizations simultaneously: the widely pursued \emph{3D photonic bandgap}.
Historically a 3D bandgap has never been considered for the propagation of X-rays in periodic media, since a gap requires a high refractive index contrast of order unity, whereas the refractive index varies by less then $10^{-4}$ in the X-ray range.
At frequencies inside the 3D photonic bandgap, the density of optical states (DOS) completely vanishes.
As the density of states can also be interpreted as the density of vacuum fluctuations~\cite{Milonni1994}, a 3D bandgap thus serves as an effective shield to these fluctuations.
The total absence of optical modes in a photonic bandgap has implications beyond classical optics that break the analogy between the behavior of light in photonic crystals and the behavior of electrons in atomic crystals.

3D photonic bandgap crystals play an important role in cavity quantum electrodynamics (cQED)~\cite{Bykov1972spj, Yablonovitch1987prl} where they offer at least five prospects for new physics\footnote{Anderson localization of light in a 3D photonic band gap~\citep{John1987PRL} is analogous to Anderson localization of electrons, see Ref.~\cite{Lagendijk2009PT}.}.
Firstly, probably the most eagerly pursued phenomenon is the complete inhibition of spontaneous emission: an excited quantum emitter - such as an atom, molecule, or quantum dot - embedded in a crystal with its transition frequency tuned to within the 3D bandgap remains forever excited since it cannot decay to the ground state by emitting a photon.
Any interaction mediated by vacuum fluctuations is affected by their suppression in the 3D bandgap~\cite{Haroche1992}.
Therefore, a crystal with a 3D photonic bandgap not only inhibits spontaneous emission - including a shift of the emitter's frequency known as the Lamb shift~\cite{John1990PRL} - it will also modify the spectrum of blackbody radiation~\cite{Babuty2013PRL}, it will affect resonant dipole-dipole interactions including the van der Waals and Casimir forces~\cite{Kurizki1988PRL, Antonoyiannakis1999PRB}, and the well-known F{\"o}rster resonant energy transfer that is prominent in biology and chemistry~\cite{Haroche1992, Novotny2006}.

A finite inhibition of spontaneous emission is also feasible in other optical materials such as microcavities or nanowires, if the emitters are carefully positioned in a tiny volume~\cite{Bayer2001PRL, Bleuse2011PRL}.
In contrast, in photonic bandgap crystals the extent over which emitters are controlled is only limited by the crystal's volume.
Moreover, the complete and radical suppression of electromagnetic modes in the bandgap is unique to photonic crystals, and is not found in other optical materials that appear to exclude all light in a particular frequency range.
For instance, if we imagine a metal-coated box, then an emitter inside such a box is not seen from the outside, and the emission might be perceived to be inhibited.
Nevertheless there are many optical states in the box in which photons can be emitted, before they are ultimately absorbed in the metal walls.
In a 3D photonic bandgap, however, there are simply \emph{no} electromagnetic states available hence an excited emitter cannot emit a photon at all and remains forever in the excited state.

Secondly, once a bandgap is achieved, the cQED physics becomes even richer by introducing point defects.
Such defects locally break the crystal's symmetry, which results in the appearance of isolated electromagnetic resonances in the bandgap.
At these resonances the field is spatially localized within a tiny nanoscale volume $V_{cav}$ that is typically less than a wavelength cubed and thus less than a cubic micron.
In other words, a point defect acts as a tiny cavity that is shielded in all three dimensions from the vacuum by the surrounding  crystal~\cite{Yablonovitch1991PRL2, Joannopoulos2008PUP}.
Hence such a photonic bandgap cavity is called a "nanobox for light".
Since the density of states in a nanobox is proportional to $1/V_{cav}$ and thus greatly enhanced by the tiny volume, an embedded emitter experiences a greatly enhanced emission rate, also known as Purcell effect~\cite{Purcell1946PR}.

A third reason why 3D photonic bandgaps are relevant to solid state cQED occurs when a gain medium is introduced in a nanobox.
Such a nanobox with gain offers the promise of a thresholdless laser~\cite{Yablonovitch1987prl}:
Since only one resonance exists in a photonic bandgap cavity, and the vacuum is shielded, there is no competing spontaneous emission into modes other than the lasing mode, and the laser thus immediately switches on~\cite{Bjork1994PRA}.
Moreover, since 3D photonic bandgap crystals are typically semiconductor devices, the on-chip integration of such a thresholdless laser is readily foreseeable.

Fourth, an important research theme in cQED is the breaking of the weak-coupling approximation.
There are in essence two ways to break this limit: The most well-known approach consists of embedding a two-level quantum emitter in a high-finesse cavity, and tuning the emitter frequency $\omega_{eg}$ to the cavity resonance frequency $\omega_{cav}$.
When a quantum of energy is exchanged between the emitter and the cavity at a rate $\Omega_{R}$ - called the vacuum Rabi frequency - that exceeds leakage rates such as the spontaneous emission rate or the cavity escape rate, the emitter and the cavity resonance are in a coherent superposition, called QED strong coupling~\cite{Haroche1992}.
While the achievement of this limit has been realized with pillar cavities~\cite{Reithmaier2004N}, ring cavities~\cite{Peter2005PRL}, and 2D photonic crystal cavities~\cite{Yoshie2004N}, the realization in nanoboxes is still outstanding.
There are several alternative ways to break the weak-coupling limit.
One approach is to operate close to a van Hove singularity where the density of states has a cusp~\cite{Ashcroft1976}.
A second approach to break the weak-coupling limit consists of rapidly modulating the "bath" that surrounds a two-level quantum emitter~\cite{Lagendijk1993Lucca}, using ultrafast all-optical switching methods~\cite{Johnson2002PRB}.

Fifth, in quantum physics there is an active interest in decoherence, that is, the loss of coherence between the components of a system that is in a quantum superposition~\cite{Zurek1991PT}.
A consequence of decoherence is that a quantum system irreversibly reverts to revealing classical behavior, which is undesirable for applications in notably quantum information processing\cite{Nielsen2000QCQI}.
In case the quantum systems are optical emitters, cQED is the relevant field.
An important component of decoherence in cQED corresponds to the escape or emission of photons, or by absorption of the photons by the environment~\cite{Mabuchi2002S}.
Hence the shielding of vacuum fluctuations by a 3D photonic bandgap offers opportunities to make optical quantum systems robust to decoherence.

In this review we summarize recent progress on the five subjects in cQED listed above, with emphasis on experimental work.
In addition, we provide an overview of the current status regarding the fabrication of 3D photonic bandgap crystals.
For the scope of this review, we limit ourselves to \emph{i)} classes of photonic crystals with demonstrated 3D photonic bandgaps, and \emph{ii)} light frequencies in the optical regime, chosen to correspond to wavelengths $\lambda \le 2500$ nm (or frequencies $\omega/(2 \pi) \geq 10^{14}$ s$^{-1}$) as these are accessible by conventional optics available in many laboratories all over the world.
Moreover, at these frequencies stimulated and spontaneous emission rates are highly significant.

\section*{Theory}
\subsection*{Spontaneous emission control}
In the weak-coupling approximation of quantum electrodynamics (QED), the radiative rate of spontaneous emission $\gamma_{rad}$ of a two-level dipole quantum emitter is given by Fermi's "golden rule"~\cite{Fermi1932RMP}.
It is well-known that spontaneous emission is not an immutable property of an emitter - such as an excited atom, molecule or quantum dot - but it is also influenced by the emitter's nearby environment~\cite{Purcell1946PR, Kleppner1981prl}.
The influence of any non-dissipative environment is described by the local density of optical states (LDOS) $N(\omega, \mathbf{r}, \mathbf{e}_d)$ that counts the number of photon modes available for emission weighted with their amplitude squared, and that is interpreted as the density of vacuum fluctuations~\cite{Sprik1996epl}.
Control of QED properties by means of confined light - that is expressed by a modified LDOS - is generally considered to be the realm of cavity QED~\cite{Haroche1992}.
The radiative rate for the transition from the excited state $|e\rangle$ to ground state $|g\rangle$ is conveniently expressed as~\cite{Vats2002PRA,Novotny2006}
\begin{equation}\label{eq:rate_LDOS}
\gamma_{rad}(\omega_{eg},\mathbf{r},\mathbf{e}_d) = \frac{\pi d^2 \omega_{eg}}{\hbar \epsilon_0} N(\omega = \omega_{eg}, \mathbf{r}, \mathbf{e}_d),
\end{equation}
with $\omega_{eg}$ the emission frequency, $\mathbf{r}$ the position of the emitter, $\mathbf{e}_d$ the dipole orientation, $d$ the modulus of the transition dipole moment matrix element, and $\hbar$ Planck's constant.
It is instructive to briefly summarize the approximations and assumptions used to derive the time-evolution of the emitter's exited state and thus Fermi's "golden rule" Eq.~(\ref{eq:rate_LDOS}):\renewcommand{\theenumi}{\alph{enumi}}
\begin{enumerate}
\item The electric-dipole approximation is applied to the emitter to evaluate the electric-field operator.
\item The perturbation expansion is used to describe the time evolution of the emitter-field system.
\item The rotating-wave approximation is used to neglect rapidly-changing counter-rotating field terms.
\item The dielectric function $\epsilon(\mathbf{r})$ is taken to be real to properly define modes and hence the LDOS (see Eq.~\ref{LRDOS} below.)
\item The emitter-field interaction is taken to be a \emph{Markovian} process, which assumes that if a photon is emitted the memory of this event is lost practically instantaneously by the quantum system, thus corresponding to a vanishing coherence time~\cite{Milonni1994,Lagendijk1993Lucca}.
\end{enumerate}

From equation~(\ref{eq:rate_LDOS}) it is apparent that one can control the emission rate by means of the LDOS.
The prefactor contains intrinsic emitter properties namely the transition dipole moment \emph{d} and transition frequency $\omega_{eg}$.
Equation~(\ref{eq:rate_LDOS}) reveals the well-known fact that the emission rate depends on the frequency and the position of the emitter.
While many efforts are directed at controlling spontaneous emission by various classes of nanophotonic systems - cavities, antennae, plasmonic systems, random media - arguably the most radical change in emission occurs in a 3D photonic bandgap where the LDOS vanishes, see Figure~\ref{fig:LDOS-freq}.
Therefore, an emitter in a 3D bandgap that is in the excited state is radically forbidden to decay by emitting a photon.
In other words, while the excited state $|e\rangle$ is usually unstable against decay to the ground state $|g\rangle$ under the emission of a photon, the excited state is stabilized in a 3D photonic bandgap!

\subsection*{The local density of optical states}

\begin{figure}[!htb]
\begin{center}
\includegraphics[width=0.9\columnwidth]{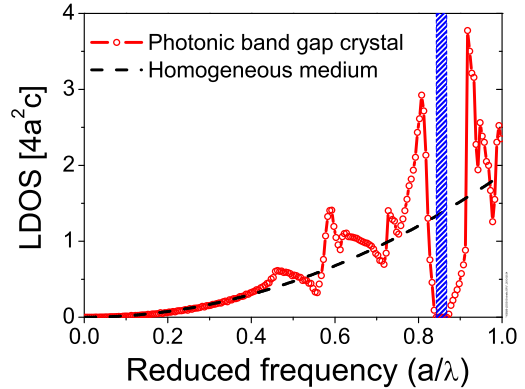}
\caption{
Local density of optical states (LDOS) versus frequency in a 3D photonic bandgap crystal.
The LDOS was calculated for an infinite inverse opal with an $\epsilon = 11.9$ backbone (red connected circles), relevant for silicon.
The LDOS was averaged over the unit cell such that it represents the total density of states. 
The blue hatched bar indicates the range where the LDOS vanishes: the 3D photonic bandgap.
The black dashed curve is the LDOS of a homogeneous medium with an effective refractive index similar to that of the crystal.
The frequency is reduced with the lattice constant $a$.
Data from Ref.~\cite{Nikolaev2009JOS}.}
\label{fig:LDOS-freq}
\end{center}
\end{figure}

The local density of optical states (LDOS) $N(\omega, \mathbf{r}, \mathbf{e}_d)$ is defined as:
\begin{eqnarray}
\label{LDOS2Green}
N(\omega,\mathbf{r},\mathbf{e}_{d}) \equiv \frac{6 \omega}{\pi c^2}
(\mathbf{e}_d^T \cdot \mathrm{Im}(\mathbf{G}(\omega;\mathbf{r},\mathbf{r})) \cdot \mathbf{e}_d),
\end{eqnarray}
%
\noindent
with $\mathbf{G}(\omega;\mathbf{r},\mathbf{r})$ the Green dyadic \cite{Novotny2006} that depends on frequency and position.
The LDOS is a classical quantity, which can be appreciated from the absence of Planck's constant from Eq.~(\ref{LDOS2Green}).
In case of dissipative optical media with complex $\epsilon$, the Green dyadic is well defined, and the imaginary part of the Green dyadic describes the \emph{total} decay rate, \emph{i.e.}, the sum of the radiative decay rate and the non-radiative quenching rate induced by the dissipative environment.
In dielectric photonic crystals, it is more tractable to calculate the LDOS by a summation over all Bloch modes as~\cite{Sprik1996epl}:
\begin{equation}\label{LRDOS}
N(\omega,\mathbf{r},\mathbf{e}_d) \equiv \frac{1}{(2\pi)^3 \epsilon(\mathbf{r})}
\sum_p\int^\infty_{-\infty}d\mathbf{k} \delta(\omega-\omega_{\mathbf{k},p})|\mathbf{e}_{d}\cdot\mathbf{\Lambda}_{\mathbf{k},p}(\mathbf{r})|^2,
\end{equation}
with $\mathbf{\Lambda}_{\mathbf{k},p}(\mathbf{r})$ a field mode with wavevector \textbf{k} and polarization state \emph{p} = 1,2.
For cavity QED it is important to note that the field modes are projected on the orientation of the transition dipole moment $\mathbf{e}_{d}$. 
Since this expression for the LDOS employs the plane-wave expansion~\cite{Joannopoulos2008PUP}, the LDOS in Eq.~(\ref{LRDOS}) pertains to an infinitely extended crystal $L \rightarrow \infty$. 
In the limit of a homogeneous dielectric medium with a spatially independent refractive index $n = \sqrt{\epsilon}$, the LDOS is equal to 
\begin{equation}\label{LRDOS_homog}
N(\omega) \equiv \frac{n \omega^2}{(3 \pi)^2 c^3},
\end{equation}
which illustrates the well-known dependence on the frequency squared. 

Figure~\ref{fig:LDOS-freq} shows the LDOS calculated for a 3D photonic bandgap crystal.
At low frequencies the LDOS increases quadratically, which illustrates the feature that if the wavelength is much greater than the lattice parameter ($\lambda \gg a$), the crystal effectively behaves as a homogeneous medium, \emph{cf.} Eq.\ref{LRDOS_homog}.
With increasing frequency modulations appear in the LDOS, as well as characteristic cusps called van Hove singularities~\cite{Ashcroft1976}.
The blue hatched bar highlights the range where the LDOS completely vanishes: the 3D photonic bandgap.
The complete vanishing of the LDOS over a finite bandwidth, irrespective of position $\mathbf{r}$ or dipole orientation $\mathbf{e}_d$, is a unique feature of 3D photonic crystal.
Whereas other systems such as 2D slab photonic crystals~\cite{Noda2007NP} or nanowires~\cite{Bleuse2011PRL} have deep pseudogaps and concomitant strong emission inhibition, the LDOS never completely vanishes.

At the edge of the photonic band gap the LDOS vanishes as $\sqrt{|\omega - \omega_{edge}|}$ as notably pointed out in Ref.~\cite{Li2001PRA}.
Hence the edge has a cusp and is thus also a van Hove singularity.
For many years, the LDOS was considered to diverge at the edge of the gap, which led to many intricate QED predictions~\cite{Lambropoulos2000RPP}.
It appears, however, that such a diverging LDOS is typical for 1D systems~\cite{John1990PRL}, but not for 3D crystals.

\subsection*{Quantum efficiency of the emitters and degree of cQED control}
In order to choose a suitable QED experiment with emitters in a photonic bandgap crystal or to correctly interpret results, it appears to be essential to consider at least one property of the emitters, namely the emission quantum efficiency~\cite{Koenderink2002prl, Koenderink2003pss}.
The emission quantum efficiency $\eta$ of one emitter is defined as the ratio of the radiative rate to the sum of the radiative and the non-radiative rate $\gamma_{nrad}$:

\begin{equation}\label{eq:quantum-efficiency}
\eta(\omega_{eg}) = \frac{\gamma_{rad}(\omega_{eg})}{\gamma_{rad}(\omega_{eg}) + \gamma_{nrad}} = \frac{\gamma_{rad}(\omega_{eg})}{\gamma_{tot}(\omega_{eg})},
\end{equation}
For the experimentally relevant case of an inhomogeneously broadened ensemble of emitters, the quantum efficiency becomes distributed; this situation is extensively discussed in Ref.~\cite{vanDriel2007PRB}.

\emph{High-efficiency quantum emitters.}
If the quantum efficiency of the emitters is high ($\eta \rightarrow 1$), the experimental method of choice is time-resolved emission.
In this method, the decay of the population density of excited emitters $N_{exc}(t)$ is probed by recording a decay curve $g(t)$.
Here $g(t)$ is the \emph{total} decay, \emph{i.e.}, the sum of the radiative decay events $f(t)$ and non-radiative decay events $(g-f)(t)$.
We are primarily interested in time-resolved emission measurements, which are generally recorded by the well-known time-correlated-single-photon-counting method~\cite{Lakowicz2008}.
The resulting decay curve $f(t)$ is the distribution of arrival times of single photons after exciting the emitters with a short laser pulse at time $t = 0$, averaged over many excitation-detection cycles.
Such a histogram is the probability density of emission which is modeled with a probability density function~\cite{Dougherty1990}.
The emission by the excited emitters at time $t$ is described by a reliability function or cumulative distribution function equal to $\left(1-\frac{N_{exc}(t)}{N_{exc}(0)}\right)$~\cite{Dougherty1990}.
The reliability function tends to 1 in the limit $t \rightarrow \infty$ and to zero in the limit $t \rightarrow 0$.
The relation between the fraction of excited emitters and the decay curve, in other words, between the reliability
function and the probability density function is

\begin{equation}\label{eq:master-emission}
\int_{0}^{t} g(t')dt' = 1 - \frac{N_{exc}(t)}{N_{exc}(0)}.
\end{equation}
Physically equation~(\ref{eq:master-emission}) means that the decrease of the density of excited emitters at time $t$ is equal to the integral of all prior decay events~\cite{vanDriel2007PRB}.
Equivalently, the total decay $g(t)$ is proportional to the time derivative of the fraction of excited emitters.
In many reports the distinction between the reliability function and the probability density function is neglected; the intensity of the decay curve $g(t)$ is assumed to be directly proportional to the density of excited emitters $N_{exc}(t)$.
This proportionality only holds for single-exponential decay but not for the general case of non-single-exponential decay; this distinction has important consequences for the interpretation of non-single-exponential decay that often occurs in photonic crystals~\cite{vanDriel2007PRB}.

From the slope of the photon arrival time distribution, we obtain the total decay rate that is equal to $\gamma_{tot} = \gamma_{rad} + \gamma_{nrad}$.
Since an efficient emitter has $\gamma_{rad} \gg \gamma_{nrad}$, a time-resolved experiment yields to a good approximation the radiative rate $\gamma_{rad}$ that one is interested in.
Since the radiative rate depends on the density of states, see Eq.~(\ref{eq:rate_LDOS}), it is generally independent of the direction of emission.
While there are special situations that the rate may depend on detection angle (see Ref.~\cite{Vos2009PRA}), one should beware of possible artifacts in this respect.

The average arrival time of the emitted photons or the average excited-state lifetime is given by the first moment of the time-resolved emission curve $f(t)$~\cite{vanDriel2007PRB}.
This result confirms that the dynamics of the population of the excited state $N_{exc}(t)$ is controlled by embedding emitters in a photonic band gap crystal.

\emph{Low-efficiency quantum emitters.}
If the quantum efficiency of the emitters is low ($\eta \ll 1$), the experimental method of choice is the continuous-wave (cw) observation of the emitted intensity~\cite{Koenderink2002prl, Koenderink2003pss}.
From rate equations one can derive that the emitted intensity $I(\omega_{eg})$ is equal to

\begin{equation}\label{eq:cw-intensity}
I(\omega_{eg}) = \gamma_{tot}(\omega_{eg}) N_{exc} \eta(\omega_{eg}) = P \eta(\omega_{eg}) = P \frac{\gamma_{rad}(\omega_{eg})}{\gamma_{tot}(\omega_{eg})},
\end{equation}
where $P$ is the rate of excitation of the emitters~\cite{Koenderink2003pss}.
If the quantum efficiency is low then $\gamma_{rad} \ll \gamma_{nrad}$, hence $\gamma_{tot} \rightarrow \gamma_{nrad}$ so that $\gamma_{tot}$ is independent of the LDOS.
Consequently the cw intensity $I(\omega_{eg})$ of the emitters becomes proportional to $\gamma_{rad}$, hence $I(\omega_{eg})$ is then an excellent probe of the LDOS.
If one studies the cw intensity $I(\omega_{eg})$ to obtain information on the crystal's LDOS, care must be taken to distinguish spectral features from several other effects, such as angular effects related to photonic bandstructures, as well as the angle-dependent excitation and the angle-dependent collection efficiency~\cite{Koenderink2003pss, Husken2013jpc}.
To interpret results in terms of the LDOS, the observed intensity spectrum $I(\omega_{eg})$ is normalized to a reference that employs the same light sources in the same chemical environment.
Hence the same transition dipole moment and the same quantum efficiency pertain, so that equations~(\ref{eq:rate_LDOS},~\ref{eq:quantum-efficiency},~\ref{eq:cw-intensity}) are properly interpreted.
Second, a reference must have a well-known LDOS.
Third, a reference is required with an environment where the emitters experience the same refractive index as in the photonic bandgap crystal, to avoid complications arising from Lorentz local field factors~\cite{Schuurmans1998PRL}.
Fourth, since cw intensity is very sensitive to the collected solid angle, it is important that both the crystal and the reference have similar escape functions for internal light escaping to the detection system, or at least very well known escape characteristics.
It is our experience that the best suited reference is a similar photonic crystal in the long wavelength limit, for several reasons: the LDOS is well-known and proportional to $\omega^2$, the average index is the same hence Lorentz local field factors cancel, and the escape function is well-known and matches a Lambertian distribution~\cite{Koenderink2003pss}. 
Finally, this choice ensures a reference made by the same fabrication process, hence the same chemistry and thus the same non-radiative rate. 
Other considered reference systems such as a random medium, or a bulk material do not share these strengths and are therefore less reliable reference systems for the interpretation of the cw intensity. 

In case of low-efficiency quantum emitters, the total decay rate measured in a time-resolved experiment hardly depends on the LDOS as it is dominated by non-radiative decay events ($(g-f)(t)$)~\cite{Koenderink2002prl}.
In this situation the dynamics of the excited-state population is hardly controlled by the photonic environment, in contrast to the case of efficient emitters whose population $N_{exc}(t)$ is truly controlled by the photonic band gap environment.

\subsection*{Beyond weak coupling}[ht]
%
\begin{figure}
\begin{center}
\includegraphics[width=0.9\columnwidth]{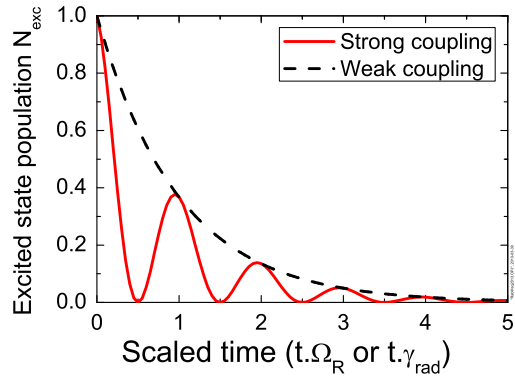}
\caption{Excited state population $N_{exc}$ of a two-level emitter with time $t$.
In the strong-coupling limit, damped vacuum Rabi oscillations appear (red full curve).
In the weak-coupling limit, the population of the excited state $|e\rangle$ decays exponentially with the spontaneous emission rate $\gamma_{rad}$ (black dashed curve).
Time is scaled with the vacuum Rabi frequency $\Omega_{R}$, or the spontaneous emission rate $\gamma_{rad}$ in case of weak coupling.
In the strong-coupling case, $\gamma_{rad}$ is taken to be equal to $\Omega_{R}$.
}
\label{fig:weakstrongcouple}
\end{center}
\end{figure}
In the Markov approximation we assume that the quasi-continuum of the field modes - the \emph{bath} - equilibrates on time scales much faster than the oscillation period of the dipole $\tau_d$.
Since the LDOS $N(\omega,\mathbf{r},\mathbf{e}_d)$ has a dimension of \emph{per frequency}, it denotes a \emph{timescale} that is usually considered to be the correlation time of the bath $\tau_b$ \cite{Lagendijk1993Lucca}.
From Eq.~\ref{LRDOS_homog} we calculate the correlation time $\tau_b$ for the LDOS at an optical frequency of $10^{15}~s^{-1}$  in a quantization volume $V=(2\pi c/\omega)^3$ to be $\tau_b = 10^{-18}$ s, which is much shorter than $\tau_d \simeq 10^{-15}$ s.
The interaction with the fast bath destroys the emitter's memory of the past, which leads to the exponential decay of the excited state.
The situation $\tau_b \ll \tau_d$ is also known as the weak-coupling limit of cavity QED~\cite{Haroche1992, Milonni1994}, and pertains to the majority of spontaneous emission studies in nanophotonics with photonic crystals.

Highly interesting quantum physics arises when the weak-coupling limit is violated.
When the LDOS is strongly increased, the relaxation of the bath becomes slower.
Once the correlation time $\tau_b$ becomes of the order of, or even longer than the dipole oscillation period $\tau_d$, the quantum system retains a memory of earlier times, hence approximation \emph{e} of the weak-coupling limit is violated.\footnote{The approximations \emph{a} through \emph{e} referred to here are the approximations listed in section "Spontaneous emission control" above.}
This means that the quantum emitter and the bath are becoming strongly coupled, and the evolution of the emitter's excited state $|e\rangle$ does not show exponential decay anymore.
In this situation - typically studied with a high-Q cavity~\cite{Reithmaier2004N,Yoshie2004N,Peter2005PRL} - the emitter resonance and the cavity resonance hybridize, such that the emitters's excited state population oscillates in time, as illustrated in Figure~\ref{fig:weakstrongcouple}.

In addition to raising the LDOS by a cavity resonance, there are other ways to violate the weak-coupling limit.
One approach is to operate close to a van Hove singularity~\cite{Ashcroft1976}.
Such a singularity manifests as a cusp in the density of states, causing the density of states to become non-analytical.
The non-analytic behavior means that approximations \emph{b} and \emph{e} of the weak-coupling approximations are violated.
As a result, a single emitter tuned to a van Hove singularity is predicted to exhibit non-exponential dynamics, including intricate time-dependent oscillations of the excited-state population that are called 'fractional decay'~\cite{John1994PRA, Vats2002PRA, Kristensen2008OL}.

Another approach to violate the weak-coupling limit is rapidly modulate the bath.
Hence approximations \emph{b} and \emph{c} of the weak-coupling limit are violated.
Rapid modulation of the bath can be achieved by the ultrafast switching of the optical properties of a 3D photonic bandgap crystal on ultrafast times scales, using all-optical methods~\cite{Johnson2002PRB}.

\section*{Ultimate tools for 3D photonic bandgap cavity QED}\label{sec:tools}
\subsection*{Requirements for 3D photonic bandgap crystals}
There are four main requirements to periodically ordered nanostructures in order to successfully function as 3D photonic bandgap crystals.
Firstly, the photonic interaction strength $S$ between the light and the nanostructures should be elevated.
In the introductory chapter, we have seen that $S$ is defined as the polarizability per volume.
The photonic interaction strength equals to a very good approximation of the relevant frequency bandwidth $\Delta\omega/\omega$ of a stop gap: $\Delta\omega/\omega = S$.
For photonic crystals composed of spherical scatterers, $S$ can be expressed analytically as~\cite{Vos2001Kre, KoenderinkPhDThesis}:
\begin{equation}\label{eq:S-in-crystal}
S = \frac{4 \pi \alpha}{V} = 3 \phi \frac{\epsilon - 1}{\epsilon + 2} g(K.R),
\end{equation}
where $\epsilon$ is the relative dielectric function of the spheres, equal to $\epsilon = m^2$ with $m$ the
ratio of the refractive indices of the spheres and of the surrounding medium: $m = n_1/n_2$.
Thus a high refractive index contrast $m$ is highly desired, which dictates the choice of the constituent materials.
Air with index $n = 1.0$ is a convenient low index material, and semiconductors such as silicon or GaAs are often the high index material of choice, with an index $n \simeq 3.5$.
\footnote{A very high photonic strength may also be realized with atoms trapped on lattice sites~\cite{vanCoevorden1997PRL}. 
At frequencies near their resonance, such as alkalis near the D-transition, atoms have a very high polarizability, concomitant with photonic strengths near $S = 1$~\cite{HardingPhDThesis}.}
In Eq.~(\ref{eq:S-in-crystal}), $ g(K.R) $ is the form factor of the scatterers (here: spheres in the Rayleigh-Gans limit) as a function of scatterer radius $R$ and modulus of the diffraction vector $K = |\mathbf{K}| = |\mathbf{k_{out}} - \mathbf{k_{in}}|$.
The form factor dictates both the shape of the scatterers, and the optimal filling fraction of the scatterers in the crystal~\cite{Vos2001Kre}.

The second main requirement is that the optical absorption of the constituent materials should be as small as possible.
This is borne out by the fact that a photonic bandgap is a multiple scattering phenomenon, hence at every chance that photons are absorbed, it's "game over".
The role of absorption has been studied via the imaginary part of the frequency $\omega''$~\cite{Krokhin1996PRB}.
Near the edge of the gap the square root singularity disappears for the weakest absorption considered.
Within the bandgap the density of states becomes non-zero and proportional to $\omega'' / \omega_{gap}$.
Hence for cavity QED in a 3D bandgap, absorption of light should be reduced as much as possible.
This requirement limits the use of semiconductors to frequencies below the electronic bandgap.
Thus photonic crystals made from silicon or GaAs are limited to wavelengths longer than $1100$ nm or $870$ nm.
Indeed, in the early days of semiconductor research, an empirical relation between refractive index and electronic bandgap $E_{g}$ was established called Moss' rule: $n^4 E_{g} = 77$~\cite{Moss1959}.
In the visible range, TiO$_{2}$ is a versatile high-index material with $n \simeq 2.7$, depending on its atomic crystal structure.
Thus the choice for a high index material limits the bandgap width (via the maximum index) and the operating frequency (via the electronic gap).

The third main requirement is the photonic crystal topology.
In Ref.~\cite{Economou1993PRB} it was found that for electromagnetic waves the network topology is more favorable for the appearance of gaps than the Cermet topology with isolated scatterers.
In the network topology, both low and the high-refractive index materials form a continuous network.
This requirement strongly limits the classes of relevant crystal structures, as for instance the important class of isolated scatterers in a medium - typical for suspension of colloidal nanoparticles - do not meet this requirement.

The fourth main requirement to photonic crystal structures is that the unavoidable statistical variations of the crystal dimensions - in other words, random disorder - must be as small as possible.
It is convenient to express random disorder in terms of relative variations of the lattice parameter $\delta a/a$, and of size variations of the constituent building blocks, such as the relative variation of sphere radii $\delta R/R$.
Random structural variations cause light to be scattered leading to the extinction of coherent beams~\cite{Koenderink2005PRB}.
In case of a high photonic strength typical of bandgap formation, the extinction length is limited to about $200$ lattice spacings in case of small structural variations of $1 \%$, typical of a high-quality structure.
We will see below that significant photonic bandgap effects have recently been observed on excited-state population, from which it is concluded that state-of-the-art structures have sufficiently high order for cavity QED. 
Therefore, we note that a 3D photonic bandgap seems to be more robust to disorder than slow light phenomena, which are also considered in the context of strong light-matter interaction~\cite{Huisman2012PRB}. 

A large variety of methods has been proposed to obtain many different types of 3D photonic crystals.
Examples of the fabrication of these structures are described in a multitude of reviews.
Refs.~\cite{Busch2007PR, Joannopoulos2008PUP} offer excellent introductions into the field of photonic crystals where the theory of photonic crystals, fabrication methods, and embedding of cavities are extensively discussed.
Ref.~\cite{Galisteo2011AM} discusses self-assembly as a tool to obtain both ordered 3D photonic crystals and random structures.
Ref.~\cite{vonFreymann2010AFM} reviews photonic crystal fabrication by direct laser writing.

This section focuses on fabrication methods for 3D photonic crystals with demonstrated bandgaps for light with wavelengths up to $2500$ nm.
A few notable classes of photonic bandgap crystals are identified, including inverse opals, and diamond-like woodpiles and inverse woodpiles, see Table~\ref{table_PC-types}.
In addition 3D photonic crystals containing optical cavities are discussed.
\begin{table}
\centering
\caption{Overview of three main classes of 3D photonic crystals.
The maximum calculated width of the bandgap is given with the relevant references.}
\begin{tabular}[t]{lll}
\hline\hline
Type of & Calculated max. &  Proposed\\
photonic crystal & width of the bandgap & in Ref.\\
\hline
Inverse air-sphere opals (\textit{fcc}) & 12\%~for~$n_\mathrm{Si}$~=~3.45 & \cite{KoenderinkPhDThesis} \\
Inverse woodpiles (diamond-like) & 25\%~for~$n_\mathrm{Si}$~=~3.6 & \cite{Ho1994SSC} \\
Woodpiles (diamond-like) & 18\%~for~$n_\mathrm{Si}$~=~3.6 & \cite{Ho1994SSC} \\
\hline\hline
\end{tabular}
\label{table_PC-types}
\end{table}

\subsection*{Optical signature of a 3D photonic bandgap}
An important aspect of 3D photonic crystal fabrication is their characterization.
Scanning electron micrographs (SEM) are often used to investigate the outside of the resulting structure and obtain a first impression whether the fabrication method was successful, see the SEM images in almost any photonic crystal paper in a materials  science journal.
To investigate the interior quality of the crystals it is possible to ``open'' the structures after their fabrication.
For instance, focused ion beam milling was employed to (destructively) observe the inner structure in Refs.~\cite{Schilling2005APL, vandenBroek2012AFM}.
A non-destructive characterization method for 3D nanostructures is for instance small-angle X-ray scattering~\cite{Wijnhoven2001CM}.
An experimental demonstration that 3D crystals exhibit a photonic bandgap, however, is more challenging and requires advanced optical methods and a careful analysis.
Structures with a 3D photonic bandgap have, by definition, the following properties: \textit{i)} stopbands with overlapping frequencies for all wave vectors and polarizations simultaneously, and \textit{ii)} a vanishing density of states.
These properties allow the presence of a bandgap to be probed experimentally.

Experimentally the widths of peaks in reflectivity or of troughs in transmission provide a good estimate of the width of a stopband.
To assess the photonic strength of the crystal, the width of the stopbands in experiments is often compared to the width of stop gaps from calculated bandstructures, see Refs.~\cite{Takahashi2009NM, Staude2010OL}.
Care must be exerted, however, as stopbands in reflectivity or transmission may also occur due to so-called silent modes.
These modes occur when incident plane waves cannot couple to a field mode inside the crystal with a peculiar spatial symmetry~\cite{Robertson1992prl, Sakoda1995prb, Joannopoulos2008PUP}.
Furthermore, deriving a conclusion based on comparing the measured stopbands with calculated band structures may be impeded by the fact that the calculated geometries potentially differ from real crystals due to inevitable unknowns in the fabrication or the characterization.
Moreover, the complication can arise that field modes only couple to a specific external polarization~\cite{Ho1990PRL}.
Thus polarization-resolved experiments are necessary to demonstrate that stopbands occur for all polarizations~\cite{Staude2011OL}.
It is also important to exclude spurious boundary effects by verifying that measured stopbands reproduce at different locations on the fabricated photonic crystal.
Consequently, for reflectivity or transmission experiments to serve as a reliable indicator for the presence of a bandgap they must be performed \textit{i)} such that all incident angles are probed (2$\pi$ sr solid angle), \textit{ii}) with resolved polarizations, \textit{iii)} on different exposed surfaces of the structure, and \textit{iv)} shown to be position independent~\cite{Huisman2011PRB}.

The main property of photonic bandgap crystals, namely a vanishing density of states, allows for a robust investigation of the bandgap.
However, it is challenging to experimentally probe the density of states.
The main approach is to embed suitable emitters in the crystals and measure their emission rates, as originally done for inverse opals with a pseudo gap but no 3D bandgap in the LDOS~\cite{Koenderink2002prl, Lodahl2004N}.
In a photonic bandgap the spontaneous emission of the emitters is inhibited, thus resulting in longer lifetimes of the excited state as described by Fermi's golden rule Eq.~(\ref{eq:rate_LDOS}).
Therefore the lifetime is an observable that directly relates to the density of states and one can argue that measuring the emission rate of embedded emitters is the most suitable experimental method to demonstrate a 3D photonic bandgap.
Relevant experiments are discussed in the next section on cavity quantum electrodynamics.

When describing how photonic crystals were obtained, in many papers the fabrication data are complemented with reflectivity and transmission spectra from one or a few directions only.
In light of the discussion above, in a strict sense such results are no experimental proof of a bandgap.
To give an overview of a wide spectrum of available fabrication procedures, we take a pragmatic approach and include results in which a bandgap was inferred from such measurements in this Review.
Table~\ref{table_3D-pc-fab} provides an overview of the fabrication methods and different types of photonic bandgap crystals discussed, including relevant references.

\subsection*{Inverse opals}
In inverse opals, spherical voids are stacked in a face-centered-cubic (fcc) structure.
These air spheres are embedded in a backbone material with a high index of refraction.
Typical fabrication procedures for these photonic crystals employ template-assisted assembly based on templates of close-packed, fcc-ordered colloidal spheres, infiltrating the templates with a high refractive index material, and subsequently removing the template material to obtain air-sphere crystals~\cite{Holland1998S, Wijnhoven1998S}.
For completely infiltrated inverse opals with a high-index volume fraction of $\phi$~=~26\% it was established that the refractive index contrast must exceed $m = 2.8$ in order to open a bandgap~\cite{Sozuer1992PRB}.
Subsequent calculations revealed that intricate incomplete filling of high-index material in the form of shells surrounding the spherical voids and connecting windows between the spheres yields a maximum possible bandgap width of $\Delta\omega/\omega = 12\%$ for \textit{fcc} silicon inverse opals~\cite{Busch1998PRE, KoenderinkPhDThesis}.

Silicon is an excellent backbone material due to its high dielectric constant of 11.9 (refractive index $n$~=~3.45) at $\lambda = 1550 nm$ and the availability of routine deposition methods.
Silicon inverse opals were first demonstrated in Refs.~\cite{Blanco2000N, Vlasov2001N}.
In both cases, the template was infiltrated with silicon using chemical vapor deposition.
The calculated expected widths of the bandgaps are 5 and 7\% respectively.
Reflectivity was used to demonstrate the photonic behaviour of the crystals.
Silicon inverse opals have been reviewed in Ref.~\cite{Tetreault2004AM}.
Inverse opals of other high-index semiconductors such as GaAs have also been pursued, see Ref.~\cite{Povey2006APL}.

As an alternative to opals, templates have also been fabricated by holographic lithography~\cite{Campbell2000N}, or by interference lithography~\cite{Ramanan2008APL} where silicon inverse opals were obtained.
Large defect channels are written in these structures by an additional 2-photon polymerization process step.
Two-photon polymerization was also used to obtain large waveguide-like structures in silicon inverse opals obtained by self-assembly~\cite{Rinne2007NP}.

While inverse opals are very popular on account of the relatively easy fabrication routes~\cite{Galisteo2011AM}, the maximum width of the bandgap is relatively narrow, and requires intricate optimization.
Since the bandgap is of a higher order in the band structures (2nd order Bragg)~\cite{Vos2000pla}, the gap is narrower and more sensitive to unavoidable disorder and fabrication deviations than for structures with a lower order band gap, which is a main disadvantage for inverse opals.
From calculations in Ref.~\cite{Li2000PRB} it is concluded that variations of air-sphere radii and lattice positions of less than $2 \%$ of the lattice parameter are already sufficient to completely close the bandgap.
These results indicate that in order to display a photonic bandgap, inverse opals must be made with an extremely high precision, which may be beyond the present state-of-the-art in nanofabrication.

\subsection*{Diamond-like photonic crystals}
In the early 1990s the possibility to use photonic crystals with diamond-like structures was described in a number of papers, see Refs.~\cite{Ho1990PRL, Ho1994SSC}.
These photonic crystals are extremely interesting because they have significant potential for wide photonic bandgaps.
Therefore, materials with relatively low refractive index contrasts of $m > 1.9$ suffice in order to obtain a photonic bandgap.
A simultaneous advantageous feature of 3D diamond-like photonic crystals is that the bandgap is robust to unavoidable fabrication deviations and random disorder, see Refs.~\cite{Sigalas1999PRB, Hillebrand2004PNAS, Woldering2009JAP} for calculations on this aspect.

In one of the earliest studies calculations showed that air-spheres arranged in a diamond structure with a refractive index contrast of 3.6 would give wide bandgaps of up to $28 \%$ bandwidth~\cite{Ho1990PRL}.
Unfortunately such diamond structures remain elusive to date.
In a 2004 review diamond-like photonic crystals and the efforts to obtain them have been reviewed~\cite{Maldovan2004NM}.
Here, several main results from that review are highlighted, as well as subsequent results.

\begin{figure}[!htb]
\begin{center}
\includegraphics[width=0.9\columnwidth]{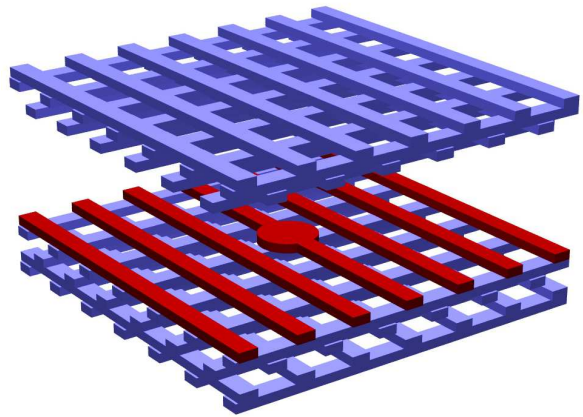}
\caption{
Schematic illustration of a 3D woodpile photonic crystal that is fabricated by a layer-by-layer approach.
The top part of the structure is lifted in order to provide a view on the central layer in which a point-defect with excess high-index material is visible.
This point-defect acts as a 3D optical cavity, or a nanobox for light.
}
\label{fig:layer-by-layer_woodpile}
\end{center}
\end{figure}

\subsection*{Woodpiles}
3D woodpile photonic crystals were originally proposed by the Iowa State group as a practical way to realize powerful diamond-structured photonic crystal~\cite{Ho1994SSC}.
The expected maximum width of the bandgap was predicted to be a sizable $18 \%$~\cite{Ho1994SSC}.
The crystal structure resembles a pile of logs of wood, hence their name.
One may argue that the analogy also pertains to the way they are often fabricated: by sequential stacking of layers of semiconductor rods, see Fig.~\ref{fig:layer-by-layer_woodpile}.
This strategy has a distinct advantage: since layers are stacked in sequential fashion it is possible to alter the layout of individual layers.
This freedom of design has been successfully used to incorporate high-quality optical cavities and waveguides in 3D photonic bandgap crystals, that are promising tools for cavity QED.
In this section a few prominent examples of woodpiles and of woodpiles with embedded optical cavities are discussed.
Alternative methods to obtain woodpiles by direct-writing and 2-photon polymerization are highlighted.

In a pioneering paper by the Sandia team, an intricate combination of several thin film deposition- and (silicon) fabrication processes was employed to obtain woodpile crystals~\cite{Fleming1999OL}.
Structures consisting of up to four layers of poly-crystalline rods were obtained.
The geometrical properties of these woodpiles were chosen such that a bandgap is expected near $\lambda = 1500$ nm.
Measured spectra indicate stop gaps with a transmission reduced to $15 \%$ at that wavelength.
While no convincing evidence for a bandgap was presented and the structures seem to have a significant misalignment between consecutive layers, kudos are in order for this pioneering work, considering that a broad stopgap with a high reflectivity has been observed~\cite{Euser2008PRB}.

The Kyoto group led by Noda has been extremely successful in fabricating woodpile photonic crystals from III-V semiconductors and from silicon using a wafer fusion technique~\cite{Noda2000S, Ogawa2004S}.
In this fabrication method, photonic crystals are fabricated by stacking semiconductor stripes, see Fig.~\ref{fig:layer-by-layer_woodpile}.
Initially, patterns of stripes - \emph{i.e.} the rods in the woodpile - are formed by electron-beam lithography and reactive ion etching in a single crystalline layer.
In case of GaAs crystals, the layer is GaAs that sits on an (Al)GaAs etch-stop layer, on a GaAs substrate.
A pair of patterned wafers is stacked and bonded in a crossed configuration with precise control over the alignment, and one of the substrates is removed.
These process steps are repeated.
At each fusing step, the number of layers doubles, until the desired structure is obtained.
One limitation of the fabrication method is that it is so advanced that it has only successfully been realized by the Kyoto group.
While initial crystals were only 4 to 8 layers thick, much thicker crystals were later realized.
Notably, impressive GaAs woodpile structures have been reported with a thickness of 17 layers of rods~\cite{Ogawa2008EL}.
To investigate the presence of the 3D photonic bandgap, normal incidence transmission and reflection spectra were collected, together with transmission spectra as a function of angle of incidence.
In the latter experiments the stopband shifts to lower frequencies (longer wavelengths) with increasing incident angles, and the shift saturates at 40 to $50^\circ$.
If the range of the 3D bandgap is defined as the range where the attenuation exceeds $80 \%$, it covers the range from $1300$ to $1550$ nm.

\begin{figure}[!htb]
\begin{center}
\includegraphics[width=0.9\columnwidth]{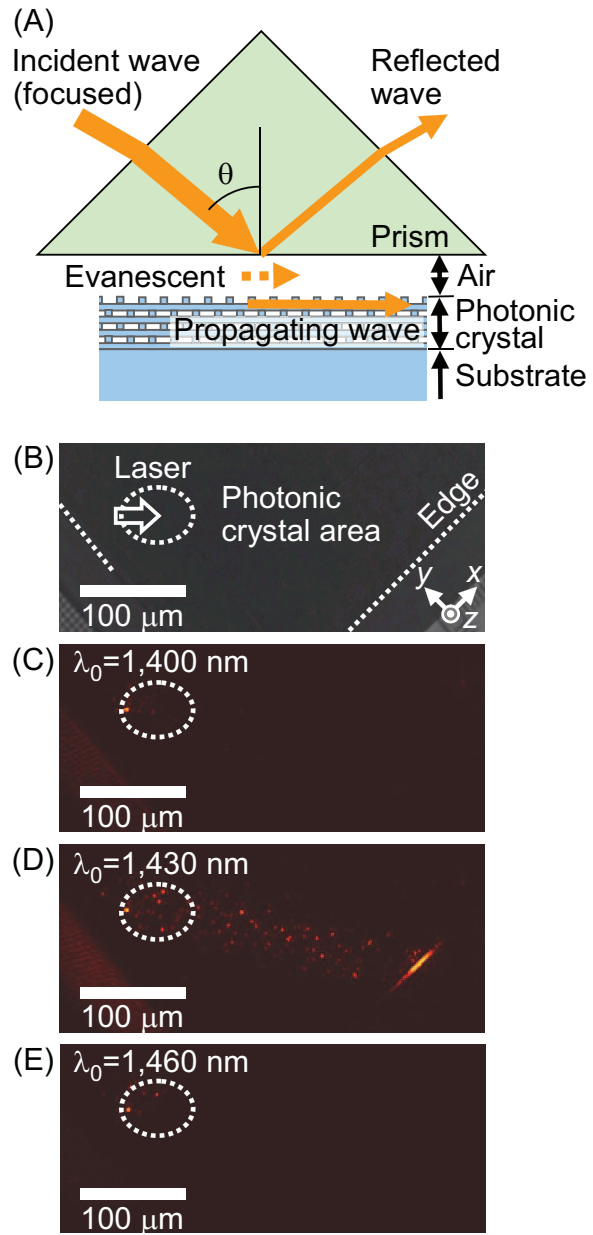}
\caption{
Experimental demonstration of surface states of a 3D photonic crystal with a 3D bandgap.
(A) Experimental set-up used to couple light into the surface modes.
(B) Optical microscope image of the surface of the 3D photonic crystal, showing the excitation point of the incident light and the edge of the photonic crystal.
(C-E) Experimental results for light propagation through the surface mode, in which the irradiating angle $\theta$ was set to $45.7 \deg$, and the wavelength to $\lambda_{0} = 1400$ nm, 1430nm, or 1460nm.
(D) Light with $\lambda_{0} = 1430$ nm propagates through the surface modes of the 3D photonic bandgap.
From Ref.~\cite{Ishizaki2009nat} with kind permission.
}
\label{fig:surface-modes}
\end{center}
\end{figure}
In an elegant experiment the Kyoto group investigated the surface modes on a 3D bandgap crystal~\cite{Ishizaki2009nat}.
It was experimentally demonstrated that photons can be confined and manipulated at the surface of 3D photonic crystals.
GaAs woodpile crystals were studied with a thickness of 8 layers and a 3D bandgap at wavelengths between $1300$ and $1600$ nm.
Since the surface states appear in the gap at frequencies below the light line, an evanescent-coupling method with a prism was used to excite the surface states, see Fig.~\ref{fig:surface-modes}(A).
Light is sent under total internal reflection conditions into the prism that is placed just above the photonic crystals.
Once the evanescent waves excite a surface mode of the photonic crystal, the reflected optical power shows a distinct minimum.
By measuring reflection spectra as a function of angle of incidence and tracking the reflectivity minimum versus wavelength, the dispersion relations of the surface states were obtained for various crystal directions.
Figures~\ref{fig:surface-modes}(B-E) show images of the photonic-crystal surface observed from the top with a camera.
Figure~\ref{fig:surface-modes}(D) clearly shows that light with a wavelength $\lambda_0 = 1430$ nm propagates along the surface of the 3D photonic crystal, at an angle of incidence $\theta = 45.7 \deg$.
The light propagates to the end of the crystal surface where it is scattered.
In contrast, at $\lambda_0 = 1400$ and $\lambda_0 = 1460$ nm light does not propagate on the surface as there are no surface modes at these wavelengths.
The wavelength of the surface modes could be tuned by scanning $\theta$.
This study is a creative demonstration of bandgap behavior, since the observations reveal that the light is strictly confined to the surface of the crystal; this is beautiful evidence that light is forbidden to propagate into the bulk of the crystal.

Ref.~\cite{Ishizaki2009nat} has also presented even more advanced manipulation of "flat light": by controlling the surface termination, the Kyoto team was able to demonstrate gaps for the surface modes.
Moreover, point defects were fabricated that behave as cavities for the surface-trapped light.
Impressive cavity quality factors up to $Q = 9000$ were reported.
This work takes an important step in opening a new route for the manipulation of light by 3D photonic crystals, as well as pioneering the surface science of 3D photonic crystals.
Moreover, the results bear intriguing analogies to surface plasmon-polaritons physics~\cite{Barnes2003N}; the absorption-free nature of a 3D photonic-crystal surface is expected to lead to new sensing applications , as well as to novel light-matter interactions that are at the heart of cavity QED.

In Refs.~\cite{Ogawa2004S, Ogawa2008EL} optical cavities were realized in woodpile crystals.
To this end, a central layer containing point defects of different sizes were incorporated in the woodpile stack.
The intensity emitted by embedded InGaAsP quantum wells served to probe the cavity quality factors.
Several peaks were observed that could be identified as cavity resonances, and a quality factor up to $Q = 350$ was measured for a thick 17-layer woodpile.
Subsequently, GaAs and silicon woodpiles were fabricated with intricate waveguide structures~\cite{Kawashima2010OE, Ishizaki2013NP}.
Guiding of light along these waveguides has been successfully demonstrated.
These examples clearly demonstrate that woodpiles show great potential for control over propagation and emission of light, including advanced functionality.

Silicon woodpiles have also been fabricated by direct laser writing of a template and subsequent double inversion to amorphous silicon~\cite{Tetreault2006AM}.
First a woodpile was fabricated from photoresist, which was fully infiltrated with silicon dioxide through chemical vapour deposition.
Subsequently, the resist template was removed to yield a silicon dioxide inverse woodpile.
The filling fraction of the inverse woodpile can be tuned by additional deposition of silicon dioxide.
In the last step, the inverse woodpile is infiltrated with amorphous silicon by chemical vapour deposition to obtain the woodpile crystal.
Based on scanning electron micrographs, the expected bandgap for this structure was predicted to have a width of nearly $9 \%$.
Optical transmission measurements displayed a strong stop gap in the range of the expected bandgap.
Innovations were reported wherein the amorphous silicon was converted to polycrystalline silicon~\cite{Staude2010OL}.
Moreover, inverse woodpiles were fabricated with embedded waveguides~\cite{Staude2011OL, Staude2012OE}.

The team from Tokyo University has successfully made woodpiles by an impressive stacking of pre-fabricated layers of semiconductor rods using a micromanipulation technique~\cite{Aoki2008NP}.
Up to 17 layers were stacked with an accuracy of 50~nm.
From bottom to top, the crystals consist of 8 GaAs layers, an active layer containing InAsSb quantum dots, and up to 8 GaAs layers.
In some cases the rods in the active layer were designed such that point defects are incorporated, see the illustration in Figure~\ref{fig:layer-by-layer_woodpile}.
For these cavity structures, quality factors up to $Q = 2300$ were reported.
It was observed that the quality factor increases by adding more top layers to the crystal.
The quality factors were improved to an impressive $Q = 38500$ by stacking a total of 25 layers and fine-tuning the size of the optical cavity~\cite{Tandaechanurat2009APL, Tandaechanurat2010NP}.

A team from MIT has made 3D silicon photonic crystals that are in essence hybrids of woodpile and air-sphere crystals~\cite{Qi2004N}.
The structures were made with an electron beam lithographic approach, where in each fabricated layer a hole section and rod section are vertically combined.
E-beam lithography allows to align each subsequent layer to the previous one with high accuracy.
Each layer is fabricated from deposited amorphous silicon and subsequent layers are deposited on top of the underlying ones. Since the lithographic pattern is defined for each layer separately, incorporation of point defects is possible.
The calculated maximum width of the bandgap for these crystals is 21\%.
A high reflectivity up to $90 \%$ was observed at wavelengths around $1300$ nm.
Troughs typical of cavities resonances were observed in the reflectivity peaks.
Unfortunately, the scanning electron micrographs reveal significant fabrication-induced structural variations in the crystals.
Nevertheless, this work is a beautiful example of layer-by-layer fabrication.

It is noted that many of the woodpile photonic crystals discussed above appear to not be cubic and thus not truly diamond-like.
This is caused by the fact that the individual stacked layers do not always possess an optimal thickness compared to the periodicity of the rods in the layers.
Hence the crystals have bandgap widths that differ from the ones calculated for true diamond-like structures.
Furthermore, layer-by-layer fabrication methods typically introduce relatively large alignment errors in the structures, which is expected to have an adverse effect on the photonic bandgap~\cite{Chutinan1999JOSAB}.

\subsection*{Inverse woodpiles}
Broad bandgaps with widths exceeding $25 \%$ have been predicted for 3D photonic crystals known as ``inverse woodpiles''~\cite{Ho1994SSC}.
These crystals consist of pores that run in two perpendicular directions in a high-index backbone.
Thus, the structure is the inverse of the woodpile structure.
Compared to woodpiles, inverse woodpiles have several advantageous features.
First, the layout and pore alignment is defined such that it is straightforward to obtain a cubic diamond-like structure with a broad gap; the cubic structure is not distorted by imperfect stacking.
Secondly, the pore diameters may be varied to optimize the volume fraction of the high index material, which is an essential tuning parameter for broad bandgaps.
In contrast, in woodpile structures an optimization of the volume fraction would entail a change of the nanorod dimensions which leads to structure distortions.

By combining macroporous etching and focused ion beam milling beautiful inverse woodpiles were first fabricated in silicon~\cite{Schilling2005APL}.
Impressive large structures were obtained although unintended misalignment reduced the width of the expected bandgap to $17 \%$.
These crystals were shown to have photonic stopbands by means of reflectivity measurements along a high-symmetry crystal axis.

Inverse woodpiles have also been fabricated by means of a sequence of \textit{i)} direct laser writing to obtain a polymer template, \textit{ii)} inversion through deposition of silicon, and \textit{iii)} removal of the template~\cite{Hermatschweiler2007AFM}.
The deposition of silicon was assisted by an intermediate silicon dioxide layer.
The resulting structures have an expected bandgap width of 14\% near a wavelength of $2500$ nm.
The occurrence of a stop gap was confirmed by means of optical reflectivity and transmission measurements.

Two-photon polymerization was used to obtain a template for a silicon inverse woodpile~\cite{Shir2010JVSTB}.
Inversion of the polymer template is achieved through a sequence of conformal coating of a layer of Al$_2$O$_3$, chemical vapor deposition of silicon, and removal of both Al$_2$O$_3$ and template.
While the demonstrated inverse woodpiles are not truly cubic and thus not diamond-like, a bandgap with a maximum width of $15\%$ was predicted for the optimal structure.
Measured reflectivity reveal a stop gap near $\lambda = 1100$ nm, with maxima up to $60 \%$.
Features in the scanning electron micrographs and absence of interference fringes in the spectra suggest that the structures made by this method suffer from imperfections such as a nonuniform surface on large length scales and roughness on smaller scales, which adversely affect bandgap formation.

In 2006 a two-directional etching method was introduced by the Kyoto group as a means to fabricate 3D inverse woodpile crystals~\cite{Takahashi2006APL}.
It was shown that cubic diamond-like structures are obtained by reactive ion etching through a suitable mask of two perpendicular sets of pores under angles of 45$^\circ$ with respect to the wafer surface.
While the structures are thin, they show strong $60 \%$ reflectivity peaks near $1600$~nm.
In 2009 a more extended studied resulted in structures displaying around $97 \%$ reflectivity~\cite{Takahashi2009NM}.
In these crystals a bandgap with a width up to 14\% is expected.
In addition, clear changes in cw intensity were observed with embedded light emitters (see next section.)
These results illustrate that these inverse woodpile are very powerful optical structures.

A similar two-directional etching method was used at Duke University to obtain woodpiles in GaAs~\cite{Tang2011IEEEJQE}.
The maximum expected width of the bandgap of optimal structures is about $18 \%$.
Reflectivity experiments reveal a strong $92 \%$ peak near a wavelength of $1300$~nm, indicative of a high photonic strength.
An outlook was presented on how to obtain microcavities at the intersection of unit cell modulations and line defects.
\begin{figure}[!htb]
\begin{center}
\includegraphics[width=0.9\columnwidth]{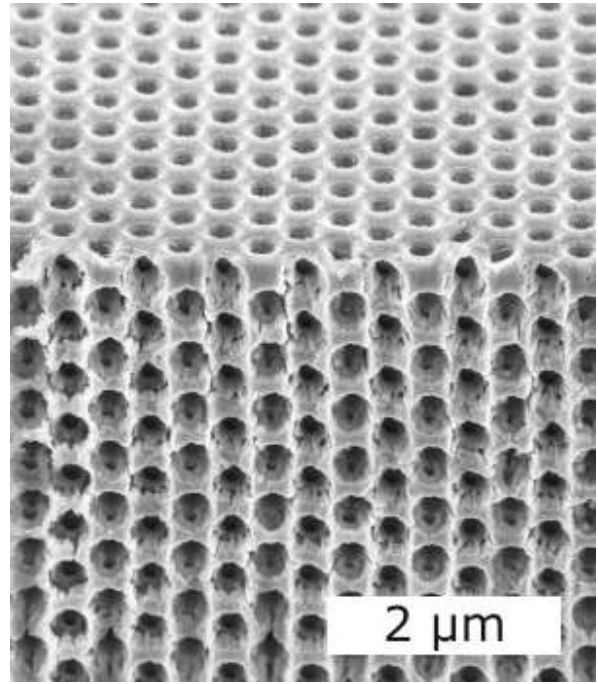}
\caption{
Scanning electron micrograph of a 3D inverse woodpile photonic bandgap crystal. 
Such a 3D nanostructure is fabricated from monocrystalline silicon using a CMOS-compatible two-directional etching technique combined with advaced alingment, see Refs.~\cite{Tjerkstra2011JVSTB, vandenBroek2012AFM}. 
This crystal has a diamond-like structure. 
}
\label{fig:inverse-woodpileSEM}
\end{center}
\end{figure}

A new two-directional etching approach using masks in inclined planes was demonstrated at the University of Twente to realize high quality diamond-like inverse woodpiles from silicon with expected bandgap widths up to $24 \%$~\cite{vandenBroek2012AFM}.
The method was developed to be CMOS-compatible in close collaboration with high-tech industrial partners.
In this method, high-purity single crystalline wafers are first etched in one direction by reactive ion etching to obtain large arrays of deep pores~\cite{Woldering2008N}.
Secondly, the sample is cleaved, rotated by $90^\circ$, and placed in a dedicated holder wafer that was developed to carefully align the samples.
Thirdly, by using the holder a second etch mask is defined in an inclined plane to the first pattern
with a high translational alignment accuracy better than $30$ nm and a high rotational accuracy better than $0.71^\circ$~\cite{Tjerkstra2011JVSTB}.
Finally, a second set of pores is etched perpendicular to the first set by deep reactive ion etching.
The overlap region of the two perpendicular sets of pores form the 3D inverse woodpile crystals, see Figure \ref{fig:inverse-woodpileSEM}.
The signature of the bandgap was demonstrated by extensive polarization-resolved optical reflectivity measurements on different crystal faces, that revealed stop bands that occur over large solid angles spanning $1.76 \pi$ sr~\cite{Huisman2011PRB}.
The maximum observed reflectivity of $65 \%$ was found to be limited by the finite crystal thickness and by surface roughness.
From the experimental observations, the bandgap was determined to have a broad relative bandwidth of $16 \%$.
We will see in the next section that these crystals are optically very powerful as they reveal prominent cavity QED effects on excited-state lifetimes~\cite{Leistikow2011PRL}.

\subsection*{Other diamond-like structures}
In 1991 a type of structures with the potential for wide bandgaps was proposed.
These structures are obtained by etching or milling pores of air sequentially in three different directions in a backbone with a high index of refraction~\cite{Yablonovitch1991PRL}.
Such crystals were later fabricated from GaAs~\cite{Cheng1996PS}.
Transmission measurements were used to study the optical properties of these structures and an attenuation up to 80\% was reported.
A width of the bandgap of around 19\% was reported, inferred from the width of the troughs in transmittance.
In addition this paper emphasizes the importance of avoiding making these structures by etching or milling of tapered pores. This fabrication deviation was shown to have a significant effect on the bandgap.

Diamond-like silicon spiral photonic crystals have been obtained by glancing angle deposition \cite{Ye2007JPDAP}.
In this method, silicon is grown by electron beam evaporation on substrates that contain tungsten seeds.
These seeds are arranged in a suitable square lattice with a lattice constant near $1000$~nm.
Spirals were grown by carefully rotating the substrate during deposition.
An optimal crystal has an expected bandgap width of nearly $15 \%$.
For the realized crystals a narrower bandgap width is expected due to a mismatch of the obtained geometry with the ideal geometry.
The photonic properties were analyzed by optical reflectivity, where peaks up to $80 \%$ were observed, centered near a wavelength $\lambda = 2000$~nm.

An intriguing method to fabricate diamond-like photonic crystals is biotemplating~\cite{Galusha2010AM}.
In this method, the diamond-like scale of a beetle is used as a template.
By a double, sol-gel based, inversion method, the template is replicated in titania.
The periodicity of these diamond-like structures is in the order of the wavelength of visible light.
The expected width of the bandgap of these structures is calculated to be around $5 \%$, but the gap is probably reduced in width or even closed due to the significant structural distortions that are apparent in the scanning electron micrographs of the structures.

In summary, we have seen that 3D photonic crystals have been fabricated by many different fabrication methods.
It appears that the potential and promise of wide bandgaps has challenged many groups all over the world to expand the boundaries of materials science and nanotechnology, resulting in novel ways to sculpt and fashion 3D nanostructures.
It is exciting to foresee how this know-how will be increasingly used to apply photonic crystals in innovations where complete control of light, including single photons, is essential.
For many of these strategies it remains a challenge to embed optical cavities.
In particular for strategies employing two-directional etching, the realization of embedded cavities must be demonstrated.
Adapting the above - or new - manufacturing strategies to this end remains an inspiring goal for future materials science research.

\begin{figure}[!htb]
\begin{center}
\includegraphics[width=0.9\columnwidth]{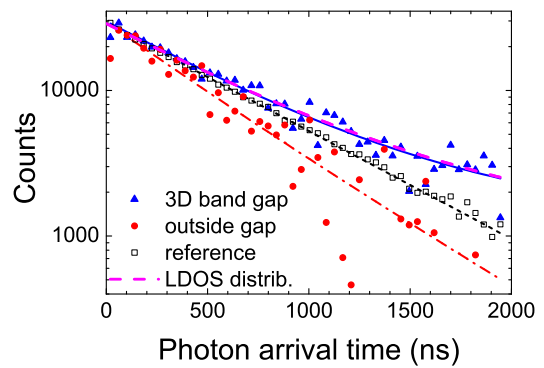}
\caption{
Distributions of photon arrival times emitted by PbS quantum dots in silicon photonic crystals with a 3D photonic bandgap (at 0.893 eV and $r = 170$ nm, blue triangles), detuned from the bandgap (at 0.850 eV and $r = 136$ nm, red circles), and in a reference (suspension, black squares).
Blue, black, and red curves are exponential models.
The magenta dashed curve is calculated from the spatial distribution of the LDOS.
The cubic crystals have lattice parameters $a = 693$ nm and $c = 488$ nm, with $a = c \sqrt{2}$, and pore radii $136 < r < 186$ nm.
Data from Ref.~\cite{Leistikow2011PRL}. }
\label{fig:decaycurve}
\end{center}
\end{figure}

\section*{Cavity quantum electrodynamics}\label{sec:cQED}

\subsection*{Inhibited spontaneous emission in a 3D photonic bandgap}
\emph{Time-resolved emission.}
Recently a first experiment was reported to study the control of the excited-state lifetime of emitters in 3D photonic bandgap crystals~\cite{Leistikow2011PRL}.
To this end, 3D inverse-woodpile crystals with a cubic diamond-like structure were made from silicon with fixed lattice parameters and a range of pore radii, so as to tune the bandgap relative to the spectrum of the emitters.
As emitters PbS colloidal quantum dots were studied at room temperature.
The dots were immersed into the crystals as a dilute suspension, hence the transition dipole orientations $\mathbf{e}_d$ sampled all directions.
The dots were kept in a toluene suspension to minimize non-radiative decay $\gamma_{nrad}$.
Although toluene as a low-index medium reduces the refractive index contrast with silicon to $m = 2.3$, inverse woodpile crystals have from the outset such a broad bandgap that the gap retained a relative width of $5 \%$.

Time-correlated single photon counting was used to precisely measure the distribution of arrival times $f(t)$ of the emitted photons~\cite{Lakowicz2008}.
Figure~\ref{fig:decaycurve} shows time-resolved spontaneous emission for quantum dots in two different photonic crystals, compared to a reference~\cite{Leistikow2011PRL}.
Emission in a crystal outside the bandgap decays faster than the reference, confirming that the excited-state lifetime of the quantum dots is controlled by the photonic crystals.
A limitation of this study is that the signal is not only emission from within the crystal, but also a background of quantum dots outside the crystal.
Thus at frequencies in the bandgap the emission is at short arrival times ($t < 500$ ns) dominated by the fast decaying background signal.
Beyond 500 ns a slow decay is apparent that corresponds to a strongly inhibited emission.
Figure~\ref{fig:decaycurve} reveals that in the bandgap the distribution of photon arrival times decays monotonically in time as predicted by Ref.~\cite{Li2001PRA}; no fractional decay or oscillations were detected as predicted elsewhere~\cite{Vats2002PRA, Wang2003PRL, Kristensen2008OL}.
Possible reasons for this discrepancy could be that the predicted features are not robust to ensemble averaging in the experiment.
Conversely, theory studies consider excited-state population dynamics instead of photon arrival times, while these two phenomena can strongly differ~\cite{vanDriel2007PRB}, or sometimes assumes an excessively large transition dipole moments.

To interpret the time-resolved spontaneous emission, the dynamics of the quantum dots in the crystal was modeled with single exponential decay, where care was taken to properly account for signal and background statistics~\cite{Leistikow2011PRL}.
Outside the bandgap the emission rate was found to be up to $2$-fold enhanced, in agreement with calculated density of states (\emph{cf.} Eq.~\ref{LRDOS}).
Within the photonic bandgap the emission rates are strongly inhibited by a factor $10 \times$ compared to emission rates outside the bandgap.
From additional observations it was concluded that the quantum dots have a very high quantum efficiency ($> 90 \%$), thus the low emission rates correspond to 10-fold enhanced excited state lifetimes, which is a clear step towards the stabilization of the excited state in a photonic bandgap.
The resulting lifetime of $T_{1} = 5.5 \mu s$ is very long for quantum dots and good news for applications in quantum information processing.

The experimental observations within the bandgap cannot be interpreted with current theories as these generally use the plane wave expansion typical of infinite crystals.
Therefore, a heuristic model was made for the spatial dependency of the LDOS in a finite photonic crystal, where it is postulated that the unit-cell averaged LDOS decreases exponentially with distance $z$ into the crystal, with a new decay length called $\ell_{LDOS}$~\cite{Leistikow2011PRL}.
This model leads to a distribution of emission rates with a minimum rate corresponding to $160 \times$ inhibition.
The corresponding time-resolved emission curve is shown in Fig.~\ref{fig:decaycurve}, where it is apparent that the curve is non-exponential due to the broad distribution, and the model agrees very well with the experiments.
The estimated LDOS decay length $\ell_{LDOS} = 1.03 a$ is $6 \times$ smaller than the Bragg length that can be derived from the photonic strength $S$ (see the Introduction chapter~\cite{Vos2015CUP}), which confirms that this is a new length scale typical for 3D band gaps.
Finally, the good agreement of model and experiments suggests that this approach is fruitful for future ab-initio cavity QED theory in 3D photonic bandgaps.

\emph{Continuous-wave intensity.}
In 2004 the Kyoto group reported the first study of emission in 3D photonic band gap crystals~\cite{Ogawa2004S}.
To this end, InGaAsP multiple quantum wells were incorporated in the central layer of 3D woodpile crystals with thicknesses between five and nine layers.
Cw emission intensity spectra $I(\omega_{eg})$ were collected at room temperature as a function of the angle of observation.
Broad stopbands were observed at constant angle, in agreement with broad stopgaps in the dispersion relation.
A strongly suppressed emission was observed in the wavelength range between 1450 and 1600 nm, independent of the angle of observation, which agrees with the expected inhibition in a 3D photonic bandgap.
Moreover, the observed range agrees well with the bandgap range identified by transmission measurements and by FDTD simulations.

Subsequently, the Kyoto group has studied emission intensity in inverse woodpile crystals~\cite{Takahashi2009NM}.
A layer of InGaAsP quantum wells was embedded in the crystal to serve as internal emitters.
By employing advanced additional wafer bonding procedures, the layer emitters was sandwiched between two halves of the photonic crystal, and carefully aligned.
Emission spectra were measured at low temperatures ($< 90$ K).
The cw intensity $I(\omega_{eg})$ was found to reduced compared to emitters embedded in bulk silicon.
While this observation represents a directional stopband, a logical next step is the demonstration of angle-independent emission inhibition.

In all studies above, the reference sample was a bulk layer of semiconductor.
Here, emitters experience a different environment and refractive index (near $3.5$) than in the photonic bandgap crystal (effective index probably $\le 2$).
Therefore it is not possible to interpret the level of inhibition, since relatively large Lorentz local field effects might come into play~\cite{Schuurmans1998PRL}.
Since all studies concern cw intensity $I(\omega_{eg})$ whose features likely correlate with the LDOS, it is concluded that the emitters must have a relatively low quantum efficiency.
This conclusion is borne out from the analysis of similar observations of inhibited cw emission in photonic crystals without a bandgap~\cite{Koenderink2002prl, Koenderink2003pss}.

Intensity studies have also been performed on various types of emitters embedded in GaAs or Si woodpile crystals made by micromanipulation by the Tokyo group.
Remarkably, no inhibited emission has been reported, see, \emph{e.g.}, Ref.~\cite{Aoki2008NP}.
It is surmised that the layer-by-layer stacking results in variations of the vertical layer spacings, which cause leaking-in of vacuum fluctuations into the crystal and thus adversely affect the inhibition.
We conclude that the role of fabrication imperfections on spontaneous emission control is currently not very well known and therefore merits further attention.

\subsection*{Emission in a nanobox for light}
An important driving force for the pursuit of 3D photonic crystals is the study of emission from a cavity in a 3D photonic bandgap that confines light in all three dimensions: a nanobox for light.
Pioneering steps in this direction were taken by the Kyoto group~\cite{Ogawa2004S, Ogawa2008EL}.
A central layer containing point defects of different sizes was incorporated in the woodpile stack.
The cw intensity emitted by embedded InGaAsP quantum wells was collected to probe the optical properties of the embedded cavity.
With increasing size of the point defect, an increasing number of resonances was identified, whose intensity has an increased baseline compared to inhibited emission in the pristine crystal.
Apparently, increasing point defect size causes an increased leakage into the photonic crystal and thus increasingly defies the shielding of the bandgap.
For the smallest point defects, a single optical resonance was observed, whose intensity had a baseline equal to the signal in the bandgap range.
The small cavities thus appear to be close realizations of a nanobox with a true 3D shielding by the bandgap.

Emission in a 3D photonic bandgap cavity has also been studied by the Tokyo group~\cite{Aoki2008NP, Tandaechanurat2009APL, Hauke2012NJP}.
In the first study, a layer of InAsSb quantum dots was embedded in the central layer where the cavity was located~\cite{Aoki2008NP}.
A strong polarization-dependent variation of the cw intensity was reported, which could be assigned to a cavity resonance with a quality factor in the range of 2000.
The main features were reproduced in a second study at low temperatures, where the quality factor was boosted to more than 8000~\cite{Tandaechanurat2009APL}.

In a third study, woodpile photonic crystals with cavities were fabricated from silicon~\cite{Hauke2012NJP}.
Ge islands were embedded, whose cw emission intensity was recorded at low temperature ($25$ K).
Two cavity resonances with narrow linewidths were observed.
The intensity of these resonances was reported to be 30 to 60 times greater compared to the reference.
Since the reference consisted of emitters in bulk semiconductor, it is very challenging to extract cavity enhancement factors (Purcell factors) from the measurements.
First, it is likely that the collection efficiency substantially differs between the photonic crystal and the bulk reference.
In bulk semiconductor, the solid angle in which emission is collected is limited by total internal reflection to a small fraction of the total solid angle.
In contrast, light in a photonic crystal is scattered over all $4 \pi$ solid angle, thus the collection efficiency is greater so that the intensity extracted from a photonic crystal may be overestimated.
Secondly, a different reference environment corresponds to a different Lorentz local field factor~\cite{Schuurmans1998PRL}.
If the reference is a semiconductor, its refractive index is much greater, leading to a much higher emission rate.
Hence, the intensity in the reference is overestimated compared to the photonic crystal.
Such effects require care to be properly accounted for.

To date, studies of emission in a nanobox concern the cw intensity $I(\omega_{eg})$ whose sharply peaked features likely correlate with the LDOS.
Therefore, the emitters have a relatively low quantum efficiency~\citep{Koenderink2002prl}.
This conclusion is confirmed by Ref.~\cite{Hauke2012NJP}, where it was reported that the Ge islands have a substantial non-radiative recombination rate $\gamma_{nrad}$.

We conclude that at this time, the challenge is open to demonstrate Purcell-enhanced emission in a 3D nanobox.
Such a demonstration requires the use of time-resolved emission, and thus the embedding of quantum emitters with an elevated quantum efficiency into a nanobox for light.

\subsection*{Laser action in 3D photonic crystal nanocavities}

The promise of miniature "thresholdless" laser action~\cite{Bjork1994PRA} in 3D photonic bandgap crystals has long been a strong motivation for the field of photonic crystals.
Important progress in this aspect has recently been reported by the Tokyo group.
GaAs woodpile crystals were fabricated with a layer of point defects that act as cavities~\cite{Tandaechanurat2010NP}.
The cavity layer was covered with an increasing number of crystal layers to increase the cavity confinement to the point of achieving a quality factor up to 38000.
The cavity layer contained highly efficient InAs quantum dots that were excited with ns laser pulses to demonstrate lasing oscillation at low temperatures.
It was observed that the peak pump power required to reach the laser threshold decreased with increasing cavity quality factors, as controlled by increasing the number of top GaAs layers.
From measurements of output power versus input power, the spontaneous emission coupling factor was determined to be $\beta = 0.54$, $0.67$, and $0.92$ for the structures with 6, 8 and 12 upper layers, respectively.
Here $\beta$ is the fraction of spontaneous emitted light that contributes to lasing; it is defined as the ratio of the power emitted into the laser mode to the total emitted power over all modes.
For the best confined structure, the value of $\beta$ was impressively close to the theoretical limit of unity for a miniature thresholdless laser.
In future, introducing a single quantum dot into the nanobox would establish an ideal solid-state system for the study of interactions between 3D confined photons and electrons enclosed in a completely controlled optical environment.

The Tokyo group has also fabricated woodpile photonic crystals with cavities~\cite{Cao2012APL}.
The crystals contained a GaAs active layer with embedded InAs quantum dots that were excited with ns laser pulses to investigate lasing at low temperatures ($11$ K).
From output power versus input power measurements, the spontaneous emission coupling factor was found to be as high as $\beta = 0.78$, which is probably a record for silicon microlasers.
The coupling factor is a little lower than in the GaAs woodpiles mentioned above, which makes sense since the quality factor is also a little lower ($Q = 22000$), hence the confinement is slightly lower.
The promise for lasing in these photonic crystals has been clearly demonstrated in these impressive experiments.
It is a thrilling prospect to see thresholdless lasing approaching room temperature, thereby opening prospects for the application of such intricate miniature lasers.

\subsection*{Ultrafast all-optical switching of 3D photonic bandgap crystals}
It is an exciting prospect in cavity QED to rapidly modulate the "bath" that surrounds a quantum emitter, and thereby enter new physical regimes~\cite{Lagendijk1993Lucca, Ma2009PRL}.
The notion to "switch" the density of states on ultrafast time scales has been first considered for 3D photonic band gap crystals~\cite{Johnson2002PRB}, where it was theoretically proposed to quickly modify the refractive index of the semiconductor backbone by exciting free carriers with short laser pulses.
As a result the 3D photonic band gap will exhibit a large shift in frequency and a change in width.
At frequencies near the gap, the LDOS may be switched from a high value to zero, or from zero to a high value, or from a high to zero to a high value on 100-fs time scales, independent of material relaxation times.
Such fast changes are expected to yield rich cavity QED behavior of excited quantum emitters.
To this end, rate equations were subsequently derived for the excited state population of two-level emitters in a time-dependent environment~\cite{Thyrrestrup2013Arxiv}.
The weak coupling approximations were used~\footnote{See Theory section.} and the LDOS was modified on time scales faster than the excited-state lifetime.
It was found that a short increase of the radiative decay rate depletes the excited state and drastically increases the emission intensity $f(t)$ during the switch event.
The time-dependent spontaneous emission revealed a distribution of photon arrival times that strongly deviated from ubiquitous exponential decay: a deterministic burst of photons will be spontaneously emitted during the ultrashort switch event.

Several experiments have been performed to study ultrafast all-optical switching of 3D photonic bandgap crystals.
In a pioneering study reflectivity changes were reported on silica opaline matrices infiltrated with Si~\cite{Mazurenko2003PRL}.
Unfortunately, this work suffered from several limitations: first, the low refractive index contrast was insufficient for a band gap.
Second, the experiments were performed at frequencies above the electronic band gap of Si, so that light is being absorbed.
Finally, the maximum feasible refractive index change was limited by the large induced absorption.
Soon after, the Karlsruhe group reported transmission changes on Si inverse opals~\cite{Becker2005APL}.
The induced absorption in their crystal was strongly reduced by annealing the Si backbone, resulting in drastically decreased Drude damping.
This frequency range in this study was limited to the range of first order Bragg diffraction where a pseudogap occurs, but no photonic band gap~\cite{Sozuer1992PRB, Busch1998PRE, Vos2000pla}.
The authors of Ref.~\cite{Euser2007JAP} performed experiments on 3D Si inverse opals in the range of the 3D photonic band gap, and similar results were obtained at Minnesota~\cite{Wei2009APL}.
Induced absorption was limited by a judicious choice of pump conditions, allowing the demonstration of a large shift of the photonic band gap~\cite{Euser2007JAP}.
Fast dynamics was observed - 500 fs up and 21 ps down - implying that switching could potentially be repeated at GHz rates.
Ultrafast switching has also been performed on Si woodpile crystals that were probed by reflectivity over an octave in frequency including the telecom range~\cite{Euser2008PRB}.
Only 300 fs after the switching pulse, the complete band gap shifted to higher frequencies before quickly relaxing within 18 ps.
The switched spectra were successfully analyzed with a theory for finite photonic crystals.

While ultrafast switching of the optical properties of 3D photonic crystals has been clearly demonstrated, the demonstration of spontaneous emission switching of emitters that experience a fast change of the LDOS is currently open.
It will be an ultimate challenge to observe the breaking of the weak coupling approximation by fast time-dependent modulation.
Relevant questions are: what is the relevant time scale for the frequency of a photon in a cavity to adjust to the shifting band gap, to what extent do such considerations apply to excited emitter states near a gap?
It will be truly exciting to see the first light shine on this subject.

\section*{Applications and prospects}
\emph{Quantum decoherence.}
There is currently a fast growing interest in quantum information science, where the goal is to store, process and transmit information encoded in inherently quantum mechanical systems~\cite{Nielsen2000QCQI}.
While many types of physical systems are being pursued, cavity QED systems involving quantum light interacting with quantum matter receive much attention.
Solid-state cavity QED offers many advantageous prospects for qubits such as system scalability and on-chip architecture, miniaturization and high speeds, and the spatial localization of the quantum emitters~\cite{OBrien2009NP}.
For the manipulation of quantum states, it is paramount to prevent decoherence, otherwise the system will behave classically~\cite{Zurek1991PT}.
It is thus desirable to increase the dephasing time $T_{2}$ of the system, which depends on the excited state lifetime $T_{1}$ and the pure dephasing $T_{deph}$ within one state: $T_{2}^{-1} = T_{1}^{-1} + T_{deph}^{-1}$~\cite{Lagendijk1993Lucca}.
We note in passing that the inverse of the dephasing time equals the linewidth that is measured in an absorption experiment.
Earlier on in this review, we have seen that the $T_1$ of quantum emitters is controlled by the LDOS, leading to already 10-fold enhanced lifetimes observed in a 3D photonic band gap.

Pure dephasing depends on both radiative and non-radiative effects.
Examples of non-radiative effects are collisions in gas phase, vibrations in liquids, or phonons in solid state.
These dephasing effects are typically controlled by cooling systems to low temperatures, or by mechanically decoupling the quantum system as much as possible from the environment by nifty spacers.
The main source of radiative dephasing is spontaneous emission from the excited state, which depends on the density of vacuum fluctuations, hence the LDOS.
Yet again, this effect is amenable to control by the nanophotonic environment.
In the extreme case of a photonic band gap, one can therefore also enhance the radiative dephasing time.
Therefore, we conclude - in agreement with elaborate theoretical calculations~\cite{Woldeyohannes1999PRA, Bellomo2008PRA} - that a 3D photonic bandgap crystal offers a favorable environment to shield qubits operating at optical frequencies from noise and fluctuations.

\emph{Resonant energy transfer.}
In cavity QED, it is a central question how multiple interacting emitters are controlled~\cite{Haroche1992, Novotny2006}.
A well-known optical emitter-emitter interaction is resonant energy transfer between pairs of dipoles where an energy quantum is transferred from one emitter, called donor, to a second emitter, called acceptor, see e.g.~\cite{Lakowicz2008}.
The involved dipole-dipole interactions are crucial to quantum information science, and (F{\"o}rster) energy transfer plays a central role in photosynthesis, as well as in photovoltaics, lighting, and molecular sensing.
Recently, the question was addressed whether energy transfer can be controlled by the photonic environment, \emph{viz.} the LDOS.
Dye molecules were separated by a short strand of double-stranded DNA, and the LDOS was controlled by positioning the FRET pairs near a mirror~\cite{Blum2012PRL}.
Contrary to early predictions, it was found that the energy transfer efficiency does change with LDOS, whereas the energy transfer rate is independent of the LDOS, in agreement with theoretical considerations.

It was predicted that in a 3D photonic bandgap the efficiency of resonance energy transfer is maximal when the bandgap is tuned to the donor emission frequency.
In case of a vanishing non-radiative decay, the efficiency will reach a perfect $100 \%$.
The observation of changing energy transfer efficiency imply a change in the characteristic F{\"o}rster distance, in contrast to common lore that this distance is fixed for a given pair of dipoles.
In case of a bandgap, this interaction distance will undergo the greatest changes.
Thus, future control of resonant energy transfer in a 3D photonic bandgap promises favorable vistas that are relevant to applications ranging from quantum information science to physical chemistry and even biophysics.

\emph{Lighting.}
Photonic crystals offer the opportunity for spontaneous emission to be strongly controlled both in spatial terms (direction of emission) or in absolute terms (rate of emission).
Therefore, it is timely to discuss the usage as practical light-emitting sources, including light emitting diodes (LED), for everyday lighting applications.
In a recent lucid review~\cite{David2012RPP}, it has been asserted that photonic crystals offer favorable strategies to efficiently couple out the light; in comparison with usually employed random surface roughening some curious common physical limitations were noted.
The Purcell enhancement of the spontaneous emission rate was also considered as a means to enhance the internal efficiency of the light sources.
It was found that such an approach is effective only for sources with specific properties, such as a small spatial extent and a narrow bandwidth so that it fits with the necessary resonant cavity.
In the review, it is also justifiably concluded that a 3D photonic bandgap is not a desired feature for lighting, as a bandgap results in the inevitable decreasing of the emission quantum efficiency (see Eq.~\ref{eq:quantum-efficiency})~\cite{David2012RPP}.
Interestingly, however, it is not widely appreciated that 3D photonic crystals reveal extended frequency ranges with enhanced density of states, which could serve to boost the emission rate and the (internal) quantum efficiency.
Calculations have revealed that silicon inverse opals have $3 \times$ enhanced emission rates over a broad $10 \%$ bandwidth~\cite{Nikolaev2009JOS}.
And provided that one controls the orientation of the transition dipole moment, a $4 \times$ enhancement is feasible over a huge octave-broad bandwidth ($\geq 100 \%$ relative bandwidth).
Therefore, we propose that the study of 3D photonic crystals at frequencies outside their bandgap may yield fruitful  applications.

In the context of lighting, it is also important to consider the important question of how to efficiently couple out of a 3D photonic band gap crystal, or how to couple light in, e.g., to excite emitters, or to address waveguides and cavities. 
In comparison to 2D slab crystals, it is not very obvious how to use incoupling gratings in 3D~\cite{David2012RPP}. 
We anticipate that the powerful new method of optical wavefont shaping - that is, the spatial addressing of incident waves with a spatial light modulator~\cite{Mosk2012NP} - will open new avenues to provide access to photonic crystals. 
In particular, based on studies where one light emitting particle in an optically thick scattering medium was selectively addressed, we are optimistic that these new techniques will for instance allow to effectively couple light into and out of a nanobox buried deep inside a photonic crystal.

\emph{Novel 3D nanofabrication strategies.}
In high-speed computing, power and especially heat dissipation issues are becoming increasingly important.
Moreover, it is being realized that continued miniaturization will meet boundaries set by fundamental physical limits~\cite{Coteus2011IJD, Ball2012N}.
Therefore in state-of-the-art circuitry in CMOS industry, an increasing number of researchers is currently contemplating to array circuits in 3D grids and networks.
A second approach to alleviate heat loads is to perform data communication on even shorter length scales by optical instead of by electrical signals.
It is our firm belief that 3D nanofabrication approaches that have been stimulated by 3D photonic crystals fabrication~\cite{Takahashi2006APL, Tjerkstra2011JVSTB} will serve as an inspiration for novel 3D CMOS fabrication strategies that are relevant for future high-speed computing~\cite{Coteus2011IJD}.

\section*{Summary}
In summary, we have seen that three-dimensional (3D) photonic crystals with a 3D photonic bandgap play a fundamental role in cavity quantum electrodynamics (QED), especially in phenomena where the local density of optical states plays a central role.
One can say that photonic bandgap crystals offer a knob to dial the density of states for broad frequency bandwidths over a wide range, from near zero to several times the value in vacuum.
We have given an overview of the current status of the fabrication of 3D photonic crystals with a bandgap at optical frequencies.
Many different methods yield powerful 3D photonic crystals.
At this time, the widely pursued woodpile crystals offer the widest versatility, as embedded high-Q cavities and waveguides have been demonstrated.
The optical experiments have been discussed that provide signatures of 3D bandgap behavior, such as broadband and wide-angle reflectivity or the observation of intricate surface modes.
We have discussed the main implications of 3D bandgaps for cavity QED, in particular spontaneous emission inhibition of emitters embedded in a 3D bandgap crystal.
Here, important progress has occurred in the last decade, bringing inhibition from a theoretical prediction to experimental reality.
We have discussed the progress in spontaneous emission and laser action of emitters placed in a photonic bandgap cavity.
Near thresholdless laser action has been observed, and its realization is now approaching room temperature operation, thereby opening avenues for applications.
The steps towards the breaking of the weak-couping limit of cavity QED have been outlined, in particular by ultrafast modulation, where experimental tools are steadily ripening.
In the final section, we have reviewed several exciting applications of 3D photonic band gaps, namely the shielding of decoherence for quantum optical systems, the manipulation of multiple coupled emitters including resonant energy transfer, lighting, and possible spin-off to 3D nanofabrication for future high-end computing.
While inhibited spontaneous emission in a 3D photonic bandgap has tested our perseverance as it took twenty-five to fourty years since the predictions by Bykov, Yablonovitch, and John~\cite{Bykov1972spj, Yablonovitch1987prl, John1990PRL}, the many recent efforts on 3D bandgaps with favorable outcomes bode well for exciting contributions to nanophotonics and beyond.

\section*{Acknowledgments}
We are grateful to all colleagues in our group for many years of pleasant and fruitful collaboration, including Ad Lagendijk, Allard Mosk, Pepijn Pinkse, Lydia Bechger, Hannie van den Broek, Tijmen Euser, Philip Harding, Alex Hartsuiker, Cock Harteveld, Simon Huisman, Bart Husken, Arie Irman, Femius Koenderink, Merel Leistikow, Peter Lodahl, Juan Galisteo Lopez, Mischa Megens, Ivan Nikolaev, Karin Overgaag, Willem Tjerkstra, Floris van Driel, Judith Wijnhoven, Elahe Yeganegi, and many, many others.
We also thank Irwan Setija (ASML), Fred Roozeboom (TNO, TUE), John Kelly and Dani\"el Vanmaekelbergh (Utrecht University), Ruud Balkenende (Philips), and many others for successful collaborations.
We are grateful to Kenji Ishizaki and Susumu Noda for their kind permission to reproduce their results.
This work is part of the research program of the ``Stichting voor Technische Wetenschappen (STW),'' and the ``Stichting voor Fundamenteel Onderzoek der Materie (FOM),'' which are financially supported by the ``Nederlandse Organisatie voor Wetenschappelijk Onderzoek (NWO)''.

\bibliography{3DpbgQEDreview}\label{refs}


\newpage
\begin{sidewaystable}
\centering
\caption{\\Overview of various types of 3D photonic bandgap crystals that have been realized.
This table provides information on the employed fabrication methods and the high-index backbone material of the crystals. Relevant references are provided in which detailed information on these structures and their fabrication can be found. For some of these references values for calculated or otherwise expected widths of the bandgaps are added between brackets. The concluding column provides additional remarks. The following abbreviations are used: 2PhP, two photon polymerization, CVD, chemical vapour deposition, DLW, direct-laser-writing, FIB, focused ion beam milling, Inv. woodpile, inverse woodpile, QD, quantum dots, RIE, reactive ion etching, SEM, scanning electron microscopy.}
\begin{tabular}[t]{lllll}
\hline\hline
Type & Fabrication method & Material & References & cQED feature \\
& & & (indication of bandgap width) & \\
\hline Inverse opal & Inversion of templates & Si &\cite{Blanco2000N}(5\%),\cite{Vlasov2001N}(7\%),\cite{Ramanan2008APL},\cite{Rinne2007NP} & Demonstrated ultrafast \\
& & & & switching~\cite{Becker2005APL, Euser2007JAP} \\
\hline \textit{Diamond-like}&&&& \\
{-Yablonovite} & RIE in 3 directions & GaAs & \cite{Cheng1996PS}(19\%) &  \\
{-Spiral crystal} & Glancing angle deposition & Si & \cite{Ye2007JPDAP}(10 to 14\%) &  \\
{-Biotemplated} & Double inversion of beetle scales & Titania & \cite{Galusha2010AM}(5\%) & Time-resolved emission in \\
& & & & crystal with pseudogap~\cite{Jorgensen2011PRL} \\
{-Inv. woodpile} & Macroporous etching + FIB & Si & \cite{Schilling2005APL}(17\%) &   \\
{-Inv. woodpile} & DLW + inversion & Si & \cite{Hermatschweiler2007AFM}(14\%) & \\
{-Inv. woodpile} & 2PhP of a template + inversion & Si & \cite{Shir2010JVSTB}(15\%) &  \\
{-Inv. woodpile} & 2-Directional etching & Si & \cite{Takahashi2006APL}(19\%),\cite{Takahashi2009NM}($>$14\%) & Embedded quantum well layer \\
{-Inv. woodpile} & 2-Directional etching & GaAs & \cite{Tang2011IEEEJQE}(18\%) & \\
{-Inv. woodpile} & 2-Directional etching with holder & Si & \cite{vandenBroek2012AFM}(24\%) & Observed inhibited and modified \\
& & & & emission of QDs in 3D bandgap~\cite{Leistikow2011PRL} \\
\hline Woodpile & Combination of many fabrication methods & Si & \cite{Fleming1999OL} & Demonstrated ultrafast \\
& & & & switching~\cite{Euser2008PRB} \\
Woodpile & Wafer fusion of layers & III-V, Si & \cite{Noda2000S}(16\%),\cite{Ogawa2004S},\cite{Ogawa2008EL},\cite{Kawashima2010OE},\cite{Ishizaki2013NP} & Incorporated cavities and quantum wells \\
Woodpile & E-beam lithography + CVD & Si & \cite{Qi2004N}(21\%) & Incorporated point defects \\
Woodpile & DLW + double inversion & Si & \cite{Tetreault2006AM}(8.6\%),\cite{Staude2010OL}(6.9\%),\cite{Staude2011OL}(15\%),\cite{Staude2012OE} & Incorporated waveguides \\
Woodpile & Micromanipulation of layers & GaAs, Si & \cite{Aoki2008NP},\cite{Tandaechanurat2009APL}(17\%),\cite{Tandaechanurat2010NP},\cite{Cao2012APL},\cite{Hauke2012NJP}(19\%) & Cavity lasing observed \\
\hline\hline
\end{tabular}
\label{table_3D-pc-fab}
\end{sidewaystable}

\end{document}